\documentclass[nofootinbib,amsmath,notitlepage,preprintnumbers]{revtex4-1}
\usepackage{multirow,mathtools}
\usepackage{amssymb,esvect,amsmath,graphicx,latexsym,amsthm,slashed,eso-pic}

         \newcommand{\cO}{{\cal O}}    
  
\newcommand{\ie}{{\it i.e.}}  \newcommand{\eg}{{\it e.g.}}
    
\newcommand{\pL}{\left(} \newcommand{\pR}{\right)} \newcommand{\bL}{\left[} \newcommand{\bR}{\right]} \newcommand{\cbL}{\left\{} \newcommand{\cbR}{\right\}}  
\newcommand{\beq}{\begin{equation}} \newcommand{\eeq}{\end{equation}}
\newcommand{\bea}{\begin{eqnarray}} \newcommand{\eea}{\end{eqnarray}}

\newcommand{\DNeff}{\Delta N_{\rm eff}}
\newcommand{\fbhi}{f_{{    \rm BH    }_i    }}

\newcommand{\lsim}{\mathrel{\hbox{\rlap{\lower.55ex\hbox{$\sim$}} \kern-.3em \raise.4ex \hbox{$<$}}}}
\newcommand{\gsim}{\mathrel{\hbox{\rlap{\lower.55ex\hbox{$\sim$}} \kern-.3em \raise.4ex \hbox{$>$}}}}

\newcommand\ddfrac[2]{\frac{\displaystyle #1}{\displaystyle #2}}

\def\lsim{\mathrel{\raise.3ex\hbox{$<$\kern-.75em\lower1ex\hbox{$\sim$}}}}
\def\gsim{\mathrel{\raise.3ex\hbox{$>$\kern-.75em\lower1ex\hbox{$\sim$}}}}

\newcommand{\Eq}[1]{Eq.~(\ref{#1})}

\newcommand{\be}{\begin{eqnarray}}
\newcommand{\ee}{\end{eqnarray}}

\newcommand{\benum}{\begin{enumerate}}
\newcommand{\eenum}{\end{enumerate}}
\newcommand{\bi}{\begin{itemize}}
\newcommand{\ei}{\end{itemize}}

\newcommand{\fbh}{f_{\rm BH}}

\begin{document}

\preprint{FERMILAB-PUB-20-125-A-T}

%\title{Gravitational Waves and Energetic Gravitons From Black Holes in the Early Universe}

%\title{Gravitational Waves and Energetic Gravitons From Early Evaporating Black Holes}
% \title{  Graviton and Gravitational Wave Backgrounds From Kerr Black Holes in the Early Universe} 

\title{ \parbox{1.2\textwidth}{Hot Gravitons and Gravitational Waves From Kerr Black Holes in the Early Universe} }

\author{Dan Hooper$^{a,b,c}$}
\thanks{ORCID: http://orcid.org/0000-0001-8837-4127}

\author{Gordan Krnjaic$^{a}$}
\thanks{ORCID: http://orcid.org/0000-0001-7420-9577}

\author{John March-Russell$^{d}$}
\thanks{ORCID: http://orcid.org/0000-0003-0483-0530}

\author{Samuel D.~McDermott$^{a}$}
\thanks{ORCID: http://orcid.org/0000-0001-5513-1938}

\author{Rudin Petrossian-Byrne$^{d}$}
\thanks{ORCID: https://orcid.org/0000-0003-0868-2950}

\affiliation{$^a$Fermi National Accelerator Laboratory, Theoretical Astrophysics Group}
\affiliation{$^b$University of Chicago, Kavli Institute for Cosmological Physics}
\affiliation{$^c$University of Chicago, Department of Astronomy and Astrophysics}
\affiliation{$^d$University of Oxford, Rudolf Peierls Centre for Theoretical Physics}

\date{\today}

\begin{abstract}

Any abundance of black holes that was present in the early universe will evolve as matter, making up an increasingly large fraction of the total energy density as space expands. This motivates us to consider scenarios in which the early universe included an era that was dominated by low-mass ($M\lsim 5\times 10^8$ g) black holes which evaporate prior to primordial nucleosynthesis. In significant regions of parameter space, these black holes will become gravitationally bound within binary systems, and undergo mergers before evaporating. Such mergers result in three potentially observable signatures. First, any black holes that have undergone one or more mergers will possess substantial angular momentum, causing their Hawking evaporation to produce significant quantities of high-energy gravitons. These products of Hawking evaporation are predicted to constitute a background of hot ($\sim$\,eV-keV) gravitons today, with an energy density corresponding to $\Delta N_{\rm eff} \sim 0.01-0.03$. Second, these mergers will produce a stochastic background of high-frequency gravitational waves. And third, the energy density of these gravitational waves can be as large as $\Delta N_{\rm eff} \sim 0.3$, depending on the length of time between the mergers and evaporation. These signals are each potentially within the reach of future measurements.

\end{abstract}

\maketitle

%%%%%%%%%%%%%%%%%%%%%%%%%%
%%%%%%%%%%%%%%%%%%%%%%%%%%
%%%%%%%%%%%%%%%%%%%%%%%%%%

% 		 Section 1: Introduction

%%%%%%%%%%%%%%%%%%%%%%%%%%
%%%%%%%%%%%%%%%%%%%%%%%%%%
%%%%%%%%%%%%%%%%%%%%%%%%%%

\section{Introduction}

Although our universe is observed to be approximately homogeneous on cosmological scales, there may exist significant inhomogeneities on scales smaller than those probed by measurements of the cosmic microwave background (CMB) or large scale structure. If sufficiently large in amplitude, such small-scale density perturbations could lead to the formation of a cosmological abundance of primordial black holes (BHs)~\cite{Carr:1974nx,Carr:1975qj}. Alternatively, primordial BHs could have been generated through bubble wall collisions following a first order thermal phase transition~\cite{Hawking:1982ga,Sasaki:1982fi,Lewicki:2019gmv} or a first order quantum phase transition~\cite{GarciaGarcia:2016xgv}, or through the process known as scalar fragmention~\cite{Cotner:2017tir,Cotner:2018vug,Cotner:2019ykd}. If these BHs were formed at or near the end of inflation, we expect them to have masses comparable to the horizon mass during that epoch~\cite{GarciaBellido:1996qt,Kawasaki:2016pql,Clesse:2016vqa,Kannike:2017bxn,Kawasaki:1997ju,Cai:2018rqf,Yoo:2018esr,Young:2015kda,Clesse:2015wea,Hsu:1990fg,La:1989za,La:1989st,La:1989pn,Weinberg:1989mp,Steinhardt:1990zx,Accetta:1989cr,Holman:1990wq,Hawking:1982ga,Khlopov:1980mg}:
\be
M_{ H} \sim \frac{1}{2GH_I} \sim 10^4 \, {\rm g} \left(\frac{     10^{10} \, {\rm GeV}  }{H_I} \right) ~,
\ee 
where $H_I$ is the Hubble rate during inflation. 
Since BHs are non-relativistic in the early universe, they redshift like matter $(\rho_{\rm BH} \propto a^{-3})$ and therefore
constitute an increasingly large fraction of the total energy density as space expands. This motivates us to consider scenarios in which BHs dominate the energy density of the early universe prior to their Hawking evaporation. BHs with $M \lsim 5\times 10^{8}$~g will evaporate before the onset of Big Bang Nucleosynthesis (BBN), and thus remain largely unconstrained by existing observations.

In previous work, we considered Hawking radiation from primordial BHs as a mechanism to produce dark matter~\cite{Lennon:2017tqq,Hooper:2019gtx} (see also, Refs.~\cite{Morrison:2018xla,Fujita:2014hha,Allahverdi:2017sks}) or dark radiation~\cite{Lennon:2017tqq,Hooper:2019gtx}.  Since Hawking radiation emits all kinematically accessible particles, regardless of their couplings, BHs represent an attractive way to generate cosmologically interesting abundances of very feebly interacting particles. In particular, the measured dark matter abundance can be produced in a BH dominated era for masses between $m_{\rm DM}\sim 10^{9}$ GeV and the Planck scale. Furthermore, if there exist any light, decoupled particle species (\eg~axions), these will also be produced as components of 
Hawking radiation and contribute to the density of dark radiation, typically reported
 in units of equivalent extra neutrino species, $\Delta N_{\rm eff}$.
 
If the early universe had ever been dominated by Schwarzschild BHs, each type of new light, decoupled particle species is robustly predicted to contribute to $\Delta N_{\rm eff}$ at the following level~\cite{Hooper:2019gtx}:  
\be
\label{neff-intro}
\hspace{2cm}~~\Delta N_{\rm eff} \simeq       \begin{cases}
0.05-0.1~~~~~~ \,\rm Real \, Scalar \\
0.1-0.3~~~~~~~ \, \, \rm Dirac \,Fermion \\
0.02-0.04 ~~~~ \, \, \rm Massless \,Vector  \\
0.07-0.14 ~~~~ \, \, \rm Massive \,Vector  \\
0.003 -0.006 ~~  \rm Graviton
\end{cases}~~    \text{ (Schwarzschild BH Domination) },
\ee
where the larger value for each species corresponds to BH masses $\sim 10^9$ g, which evaporate just 
before BBN and the smaller value corresponds to 
BH masses $< 10^5$ g, whose Hawking evaporation reheats the universe to $T \gg 200$ GeV, sufficient
to produce all known particle species. Intriguingly, with the exception of the graviton, every contribution in 
Eq.~(\ref{neff-intro}) is within the projected sensitivity of stage IV CMB experiments, $\Delta N_{\rm eff} \sim 0.02$ \cite{Abazajian:2016yjj}. A major result of our paper is the generalization of this result to the case of rotating BHs, as summarized in Fig.~\ref{HawkingNeff}. 

\begin{figure}[t]
\hspace{1cm}
\includegraphics[width=\textwidth]{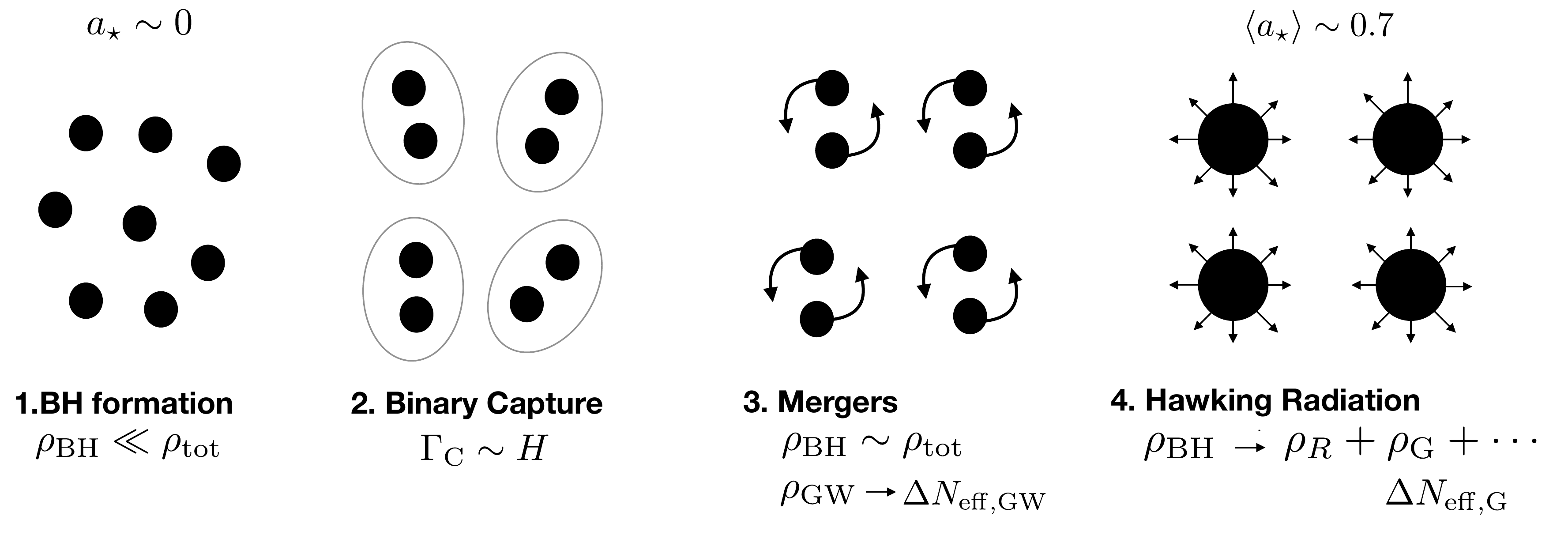} 
\caption{The time sequence of early universe events considered in this paper. {\bf (1) BH Formation:} Small-scale density
perturbations collapse after inflation to form a primordial BH population with little angular momentum, $a_\star \sim 0$. Even though the energy
density associated with this population may have been small compared to the total at this time, the 
relative fraction of the energy density in BHs increases until they evaporate. {\bf (2) Capture:} If the binary capture rate is larger than the Hubble rate, $\Gamma_{\rm bc} > H$, the BHs will efficiently form gravitational bound states. 
 {\bf (3) Mergers:} Once the capture rate freezes out, $\Gamma_{\rm bc} \sim H$, bound objects are no longer disrupted
by multi-body dynamics and can begin to inspiral, leading to the production of a stochastic background of 
high-frequency gravitational waves. The energy density in these gravitational waves can contribute significantly to $\Delta N_{\rm eff}$.
 {\bf (4) Evaporation: }
 If the BHs merge before evaporating, the population acquires significant angular momentum, $a_\star \sim 0.7$,
 which increases the proportion of Hawking radiation in gravitons. When the BH population evaporates,
 it produces Standard Model particles which thermalize to create the initial conditions for the hot radiation dominated 
 early universe. The gravitons produced as part of this radiation do not thermalize, but instead contribute to 
 $\Delta N_{\rm eff}$. 
 }
\label{cartoon}
\end{figure}

It has long been known that Hawking radiation rates are sensitive to the angular momentum of the BH~\cite{Page:1976ki}, and in this paper we revisit graviton and gravitational wave signatures that arise
 from an early universe population of evaporating Kerr BHs~\cite{Kerr:1963ud}.
Although the simplest scenarios for primordial BH generation yield an initial population of Schwarzschild BHs,
we identify regions of parameter space in which the BHs undergo one or more mergers in the early universe, resulting in a secondary BH population with substantial spin. This possibility leads to at least three potentially observable signals: 
\begin{enumerate}
%%%%%%%%
\item {\bf Gravitational Waves, $\mathbf \Omega_{\rm GW}$} \\
BH mergers in the early universe could produce a significant energy density of gravitational waves, although with a spectrum that peaks well above the range probed by detectors such as LIGO, VIRGO, BBO, ET or LISA. If these mergers occur only shortly before the BHs evaporate, future space-based gravitational wave detectors could potentially probe this signal.

%%%%%%%%
\item{  \bf Dark Radiation from Mergers, $\mathbf \Delta N_{\rm eff, GW}$ }\\
The gravitational waves generated by these BH mergers could also contribute significantly to the energy density in radiation (\ie~to $N_{\rm eff}$). If these mergers occur only shortly before the BHs evaporate, we find that $\Delta N_{\rm eff}$ can be as large as $\sim 0.3$, well within the projected reach of upcoming CMB experiments. Such a contribution could also potentially help to relax the reported Hubble tension~\cite{Riess:2019cxk,Riess:2018byc,Riess:2016jrr}. 
 %%%%%%%%
 \item{\bf Dark Radiation from Hot Gravitons, $\mathbf \Delta N_{\rm eff, G}$  } \\
 As rapidly spinning BHs preferentially radiate particle species with high spin~\cite{Page:1976ki}, this scenario can lead to the production of a significant background of energetic ($\sim$ eV-keV) gravitons, representing another potentially observable and qualitatively distinct contribution to $N_{\rm eff}$. 
\end{enumerate}

The remainder of this paper is structured as follows.  In Sec.~\ref{Sec:HE}, we discuss the process of Hawking evaporation, considering both rotating and non-rotating BHs. We find that BHs with appreciable angular momentum radiate a much larger fraction of their mass into particles with high spin, especially gravitons. In Sec.~\ref{Sec:mergers}, we calculate the timescales for BHs to form binary systems and to inspiral. We then compare this to the time required for Hawking evaporation to occur. In Sec.~\ref{Sec:NeffGW}, we calculate the energy density in the form of gravitational waves that is produced in the mergers of these BHs. In scenarios in which the BHs merge only shortly before evaporating, these gravitational waves can contribute to $N_{\rm eff}$ at a level that is within the reach of next-generation CMB experiments. In Sec.~\ref{Sec:NeffHE}, we calculate the energy density of the energetic gravitons that are produced through Hawking evaporation in this class of scenarios. Unlike in the case of non-rotating BHs, we find that the gravitons from a population of rapidly spinning BHs (such as those that have recently undergone mergers) can contribute appreciably to $N_{\rm eff}$. In Sec.~\ref{Sec:GW}, we calculate the spectrum of gravitational waves from these BH mergers and comment on the prospects for the detection of this signal. Finally, we summarize our results and conclusions in Sec.~\ref{Sec:conclusions}.

%%%%%%%%%%%%%%%%%%%%%%%%%%
%%%%%%%%%%%%%%%%%%%%%%%%%%
%%%%%%%%%%%%%%%%%%%%%%%%%%

% 		 Section 2: Evaporation 

%%%%%%%%%%%%%%%%%%%%%%%%%%
%%%%%%%%%%%%%%%%%%%%%%%%%%
%%%%%%%%%%%%%%%%%%%%%%%%%%
 
\section{Hawking Evaporation of Rotating Black Holes}
\label{Sec:HE}

Back holes emit particles and lose mass through the process of Hawking evaporation~\cite{Hawking:1974sw}. This mass loss occurs at a rate given by: 
\begin{eqnarray}
\label{rate}
\frac{dM}{dt} &=& -\ell(M,a_{\star}) \, \frac{M_{\rm P}^4}{M^2},
\end{eqnarray}
where $M_{\rm P} = 1.22 \times 10^{19}$ GeV is the Planck mass and $M$ is the mass of the BH. The dimensionless quantity, $\ell$, receives contributions from all of the particle degrees-of-freedom that are light enough to be radiated from the BH (\ie~lighter than $T_{\rm BH} = M^2_{\rm P}/8 \pi M$). Since we will only consider black holes with $T_{\rm BH} \gg 1$ TeV, $\ell$ will only depend on the angular momentum of the BH, $J$, which is related to the dimensionless spin parameter as follows: $a_{\star} \equiv J M^2_{\rm P}/M^2$ ~\cite{Page:1976ki}.

%%%%%%%%%%%%%%%%%%%%%%%%%%
%%%%%%%%%%%%%%%%%%%%%%%%%%

% 		DOF Figure < f > and < g >

%%%%%%%%%%%%%%%%%%%%%%%%%%
%%%%%%%%%%%%%%%%%%%%%%%%%%

%%%%%%%%%%%%%%%%%%%%%%%%%%
%%%%%%%%%%%%%%%%%%%%%%%%%%
As a rotating BH evaporates, it also loses angular momentum:
\be
\label{ratej}
\frac{dJ}{dt} =-h(M,a_{\star})  J \,\frac{M_{\rm P}^4}{M^3},
\ee
which can be rewritten in terms of $a_{\star}$ as follows:
\begin{eqnarray}
\frac{da_{\star}}{dt} = -a_{\star}  \frac{M^4_{\rm P}}{M^3} \biggl[h(M,a_{\star})-2 \ell(M,a_{\star})\biggr].
\end{eqnarray}
Numerical values for $h$ and $\ell$ are tabulated in Ref.~\cite{Page:1976ki} per spin degree-of-freedom (see also Ref.~\cite{Taylor:1998dk} for the scalar case).\footnote{Note the different notation in both references, $\ell \rightarrow f$ and $h \rightarrow g$.} 

%For spin $s>0$, $h_s$ is always larger than $2l_s$, so the dimensionless spin of a BH always decreases with time; not so for the scalar case $s=0$, for which an initially small $a_{\star}$ increases.

\begin{figure}[t]
\includegraphics[width=0.45\textwidth]{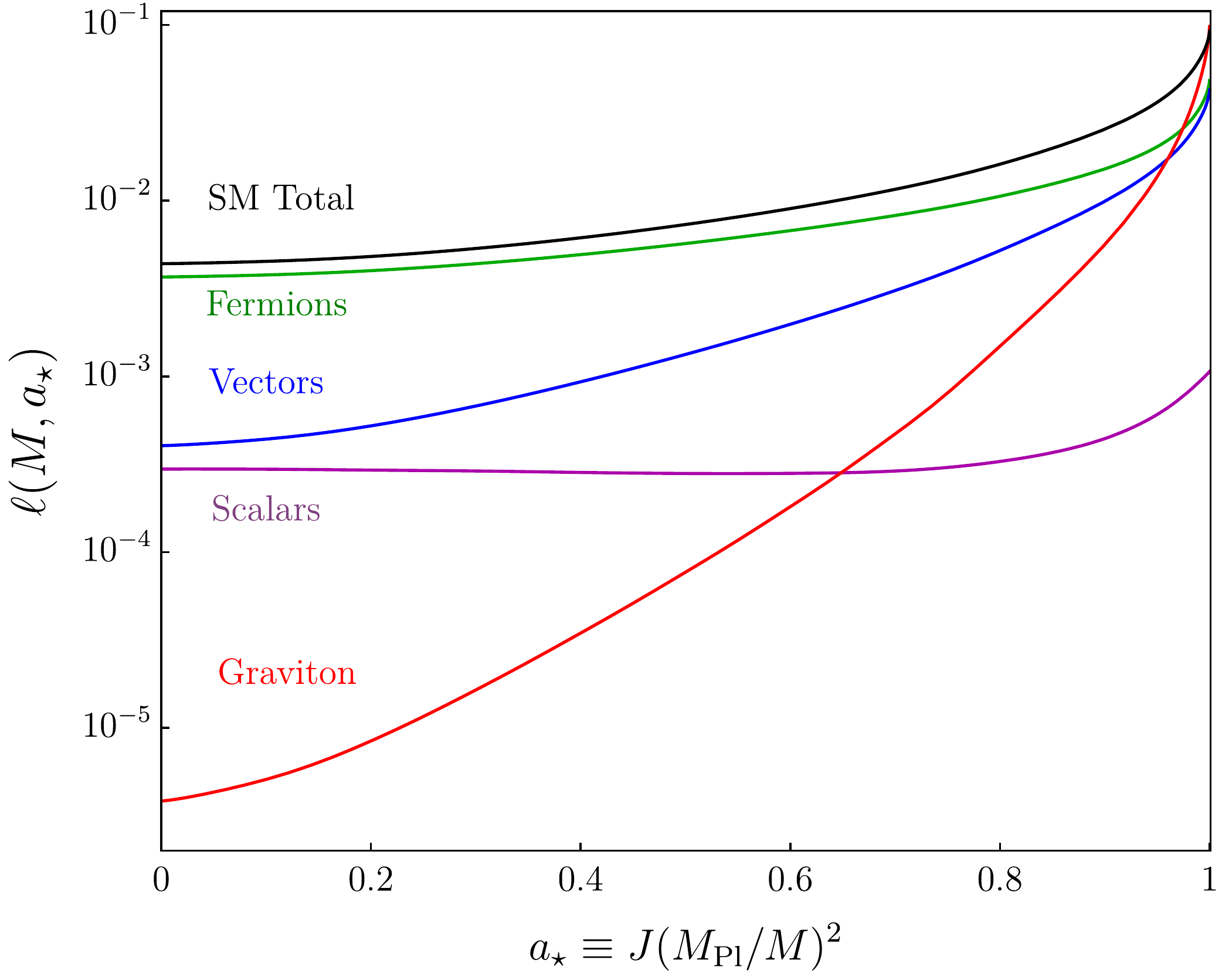} ~~~~~~
\includegraphics[width=0.45\textwidth]{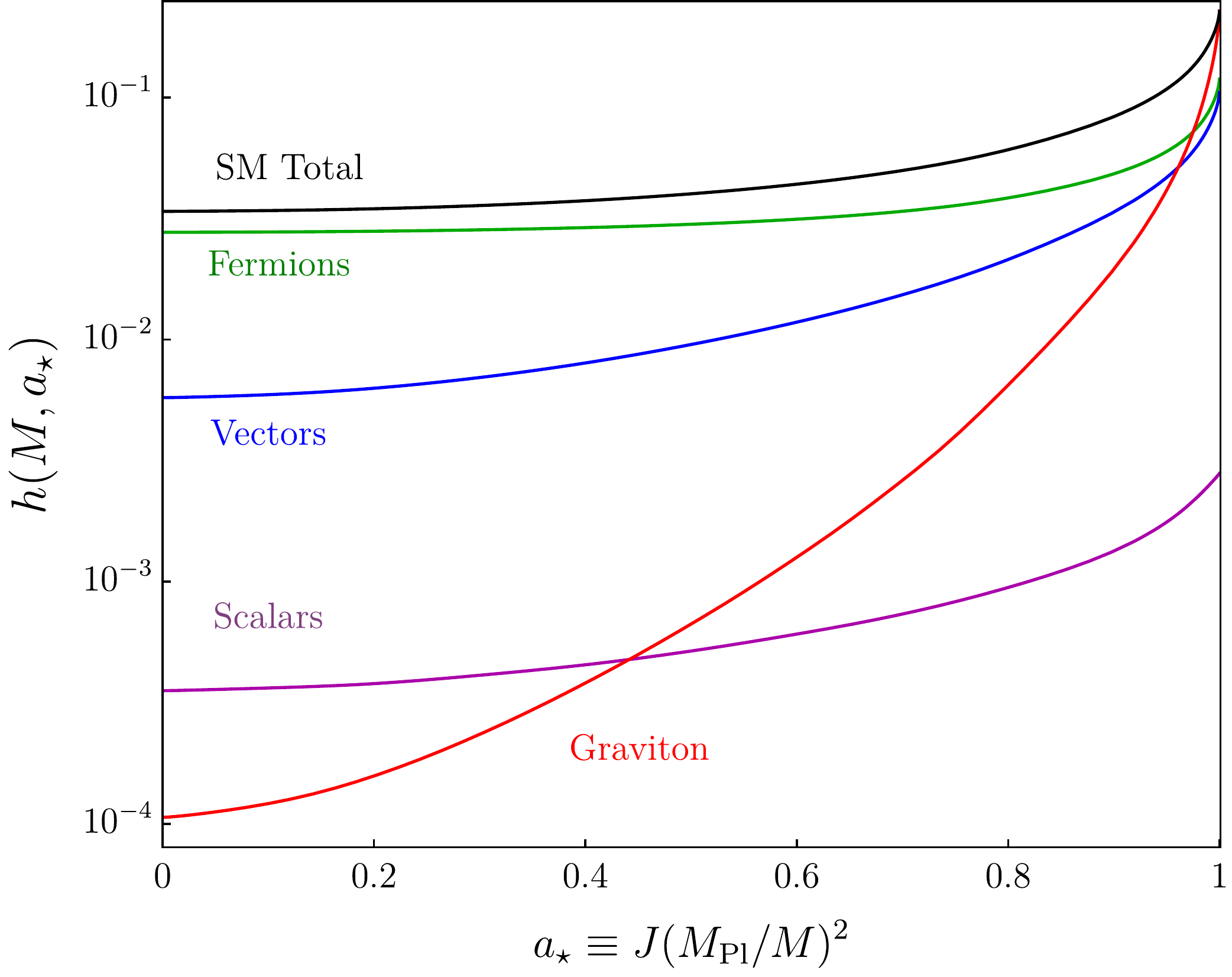} 
\caption{Values for the quantities $\ell$ and $h$ as introduced in Eqns.~(\ref{rate}) and~(\ref{ratej}), respectively, assuming the Standard Model particle content plus gravitons. All degrees-of-freedom are always active as we only consider the case $M \lsim 10^{10}$\,g (corresponding to $T_{\rm BH} \gsim$1 TeV). Also shown are the contributions to these quantities from all particles of various spins.}
\label{fgfig}
\end{figure}

Throughout this paper, we limit ourselves to the Standard Model (SM) particle content (in addition to gravitons). In Fig.~\ref{fgfig}, we plot the quantities $\ell$ and $h$ as a function of $a_{\star}$, and show the contributions from scalars, fermions, vectors, and gravitons. As a consequence of angular momentum conservation, the spin of a BH influences the rates at which various particle species are produced through Hawking evaporation. In particular, more rapidly spinning BHs radiate high-spin particles much more efficiently, including vector bosons and gravitons. 

Solving Eqns.~\ref{rate} and~\ref{ratej} numerically, we find that a BH with an initial mass, $M_i$, will evaporate entirely\footnote{We assume stable Planckian-mass BH relics either do not exist or that their presence only negligibly affects
the conclusions here.} 
in a characteristic evaporation time:
\begin{eqnarray}
\label{tevap}
\tau = \int^{M_i}_0 \frac{dM M^2}{\ell(M,a_{\star}) M^4_{\rm P}} \approx
\langle \ell^{-1} \rangle \frac{M_i^3 }{3 M^4_{\rm P}} 
=
 4 \times 10^{-4} \, {\rm s} \, \bigg(\frac{M_i}{10^8 \, {\rm g}}\bigg)^3 \bigg(\frac{\langle \ell^{-1}\rangle}{235} \bigg),
\end{eqnarray}
where $\langle \ell^{-1} \rangle$ is defined as the mass-squared weighted value of $\ell^{-1}$ from Eq.~(\ref{rate})
\begin{equation}
\langle \ell^{-1} \rangle \equiv \ddfrac{\int^{M_i}_0  \frac{  dM M^2}{\ell (M,a_\star)}   }{\int^{M_i}_0 M^2 dM}.
\end{equation}
For a non-rotating BH in the mass range of interest here, $\langle \ell^{-1} \rangle \approx 235$, whereas for a BH with maximal initial spin ($a_{\star}=1$), this quantity is approximately 35\% smaller.

In this study, we will consider primarily the case of BHs that are spinning as a consequence of having undergone previous mergers. The distribution of angular momenta predicted for such a BH population peaks strongly at $a_{\star} \sim 0.7$, almost entirely independently of the masses or the initial spin distribution of the merging binaries~\cite{Fishbach:2017dwv} (see also, Refs.~\cite{Buonanno:2007sv,Kesden:2008ga,Tichy:2008du,Healy:2014yta,Hofmann:2016yih,Gerosa:2017kvu}). With this in mind, we will adopt $\langle \ell^{-1} \rangle \simeq 195$ for BHs that have already undergone one or more mergers, corresponding to an appropriate evaporation time of $\tau = 3.3 \times 10^{-4} \, {\rm s} \, (M_i/10^8 \, {\rm g})^3$. 

Furthermore, after integrating over the evolution of BHs with a distribution of initial spins as described in Ref.~\cite{Fishbach:2017dwv}, we find that approximately $f_{\rm G}\approx 0.47\%$ of the energy emitted as Hawking radiation is in the form of gravitons (we adopt this value in our calculation in Sec.~\ref{Sec:NeffHE}). This is between three and four times as large as that found in the case of non-rotating BHs. Other formation scenarios may generate BH populations with even greater angular momenta. For example, BHs formed by the collapse of density perturbations during a matter dominated era can naturally have $a_{\star}$ very close to extremal~\cite{Harada:2017fjm,Kokubu:2018fxy}. 
For a distribution of spins peaked at $a_{\star} \sim 0.9$ ($\sim 0.95$), fractions as large as $f_{\rm G} \sim 2\%$ ($\sim 3\%$) would be appropriate. We leave a more detailed study of these scenarios to future work.

%%%%%%%%%%%%%%%%%%%%%%%%%%
%%%%%%%%%%%%%%%%%%%%%%%%%%
%%%%%%%%%%%%%%%%%%%%%%%%%%

% 		 Section 3 : Mergers 

%%%%%%%%%%%%%%%%%%%%%%%%%%
%%%%%%%%%%%%%%%%%%%%%%%%%%
%%%%%%%%%%%%%%%%%%%%%%%%%%

  \section{Black Hole Mergers in the Early Universe}
  \label{Sec:mergers}

We now turn our attention to cosmological scenarios that could plausibly result in the production of an appreciable population of Kerr BHs in the early universe. 
Although it is well known that a modified power spectrum of cosmological fluctuations on small scales can seed primordial BH formation \cite{GarciaBellido:1996qt,Kawasaki:2016pql,Clesse:2016vqa,Kannike:2017bxn,Kawasaki:1997ju,Cai:2018rqf,Yoo:2018esr,Young:2015kda,Clesse:2015wea,Hsu:1990fg,La:1989za,La:1989st,La:1989pn,Weinberg:1989mp,Steinhardt:1990zx,Accetta:1989cr,Holman:1990wq,Hawking:1982ga,Khlopov:1980mg}, this process is typically modeled as the spherically symmetric collapse of the
$\sim O(1)$ perturbations of modes that re-enter the horizon after inflation. Consequently, 
the BHs in the resulting ensemble acquire little angular momentum in the process, making a population of Kerr BHs in the early universe appear implausible. However, with an appropriate distribution of 
BH relative velocities and nearest neighbor separation distances, it is possible for the typical BH to undergo
a merger before evaporating, producing gravitational waves and increasing their angular momenta to $a_\star \sim 0.7$.  In this section, in order to identify the parameter space in which such mergers are likely, we set up the problem and highlight the relevant timescales
that govern the capture and merger rates as depicted in Fig.~\ref{cartoon}:

\begin{enumerate}
\item{\bf BH Formation:} An initial population of Schwarzschild BHs with $a_\star \sim 0$ is present in the early universe. Even
if the energy density fraction of this population starts off very small, it can grow over time to dominate the cosmic energy density
at later times.
\item{\bf Binary Capture: } If the BH binary capture rate exceeds the rate of Hubble expansion, complex multi-body dynamics can
govern the evolution of this population. However, as the universe expands, this rate eventually freezes out and the 
characteristic BH separation distance at this time sets the initial value for their subsequent evolution as binaries. 
\item{\bf Mergers:} As the binaries inspiral toward each other, they emit gravitational waves. Such signals 
may be observable if the merger timescale is comparable to the BH evaporation time. 
\item{\bf Evaporation:} Finally, as a consequence of a merger, the resulting BH possesses a typical angular momentum of $a_\star \sim 0.7$,
impacting the evaporation factors $\ell(M,a_\star)$ and $h(M,a_\star)$, as discussed in the previous section. 
The ensuing Kerr Hawking radiation now contains a much larger fraction of gravitons and other higher spin particles
relative to in the Schwarzschild scenario. 
\end{enumerate}

%%%%%%%%%%%%%%%%%%%%%%%%%%
% 	Subsection:  Setup and Assumptions 
%%%%%%%%%%%%%%%%%%%%%%%%%%
  
  \subsection{ Setup and Assumptions}

 Our setup postulates an abundance of early-evaporating primordial BHs formed at some point after inflation, but before BBN. For simplicity, 
  we will assume that this population has a common mass. For arbitrary initial conditions, the Hubble rate satisfies: 
  \be
  \label{hubble}
  H^2 \equiv \left(\frac{\dot a}{a}\right)^2 =  \frac{8\pi }{3} G  \rho_{\rm T} ~~,~~
\rho_{\rm T}(a) \equiv   \frac{ \rho_{R_i} }{a^4} +\frac{ \rho_{{\rm BH}_i}  }{a^3}  ,
  \ee
where $\rho_{\rm T}$ is the total energy density of the universe, $\rho_R $ is the density in radiation, $\rho_{\rm BH}$ is the  density in BHs, and $a$ is the cosmic scale factor, defined such that it is equal to one at the initial condition (\ie~when $\rho_{\rm R} = \rho_{R_i}$ and $\rho_{\rm BH} = \rho_{BH_i}$). Throughout this study, we adopt natural units and we define the Planck mass such that $G \equiv M_{\rm P}^{-2}$.

  We begin by defining the time-dependent quantity:
  \be
  f_{{\rm BH}} (a) \equiv  
\frac{         \rho_{{\rm BH}}(a)  }{\rho_{R}(a)  + \rho_{{\rm BH}}(a)  }.
  \ee
 If the post-inflationary universe was initially dominated by radiation at a temperature, $T_i$, Eq.~(\ref{hubble}) can be expressed as follows: 
\be
\label{HTeff}
H 
= 
1.66 \sqrt{g_\star} \frac{T_{i}^2}{M_{\rm P}} \frac{  \left(1 + \fbhi  a   \right)^{1/2}  }{a^2} 
\equiv 1.66 \sqrt{g_\star} \frac{T_{\rm eff}(a)^2}{M_{\rm P}} ,
\ee
where $g_\star$ is the number of relativistic degrees-of-freedom evaluated at $T_{\rm eff}$,
and we have defined the effective temperature, $T_{\rm eff}$, in terms of the total energy density:
\be
\label{Teff}
\rho_{\rm T} \equiv \frac{\pi^2 g_\star}{30} T_{\rm eff}^4~~~,~~~~
T_{\rm eff}(a)= 
  \dfrac{T_i}{a}  \left( 1 + \fbhi  a  \right)^{1/4} .
\ee
At the onset of BH domination, the scale factor is $\fbhi^{-1}$ and the temperature of the radiation is given by $f_{\rm BH_i}T_i$. For a sufficiently high initial temperature, the radiation density redshifts by
many orders of magnitude before BBN, so even a small initial BH fraction 
can come to dominate the energy budget \cite{Hooper:2019gtx}:
\be
\hspace{2cm} f_{{\rm BH}_i}   \gtrsim 4\times 10^{-6} \left( \frac{10^{10} \rm GeV}{T_i} \right)
 \left( \frac{10^{4} \rm g}{M_i} \right)^{3/2}
 ~~~~  (\text{Eventual BH Domination}).
\ee
This demonstrates that a BH-dominated era in the early universe is a well-motivated possibility, which could be realized across a wide range of 
post-inflationary initial conditions. 

Since we currently have no empirical access to the early universe before BBN, however, 
we will remain agnostic about the origin of this primordial BH population
and about its initial energy density. We therefore present our subsequent results in two
complementary ways: 
\begin{itemize}
%%%%%%%%%%
\item{\bf Initial Radiation Domination:} In this treatment, as outlined above, we 
assume that the post-inflationary universe is initially dominated by radiation with a subdominant initial fraction of BHs. 
In this formulation, we present $\Delta N_{\rm eff}$ results in terms of $M$, $\fbh$, and $T_{\rm eff}$
from Eq.~(\ref{Teff}). 
%%%%%%%%%%
\item{\bf Initial Black Hole Domination:}
As an alternative, we also adopt a more agnostic treatment of the early universe, in which we calculate $\Delta N_{\rm eff}$ only in terms of the BH population mass and the Hubble rate at BH domination, independent of
 how such a condition was realized. 
\end{itemize}

Finally, we note that in order to realize the interesting merger history that we consider below, we must require that the initial energy density (or equivalently the initial temperature) of early universe was quite high; occasionally higher than the nominal upper limits from CMB tensor modes, assuming single-field slow-roll inflation and other standard cosmological assumptions. However,
our scenario violates these assumptions by definition since, 1) the primordial power spectrum must be modified
relative to the predictions of single-field, slow-roll inflation and, 2) the post-inflationary evolution of a BH dominated 
early universe is nonstandard by definition. While such modifications are, in principle, still subject to some constraints
from the limits on CMB tensor modes, the precise nature of such a limit is highly model dependent and requires
a dedicated analysis; however, such questions are beyond the scope of the present work.

%%%%%%%%%%%%%%%%%%%%%%%%%%
% 		 Subsection: Binary Capture
%%%%%%%%%%%%%%%%%%%%%%%%%%
  
\subsection{Binary Capture}  
In order for a pair of BHs to undergo a merger, they must first become gravitationally bound to each other and then inspiral through the emission of gravitational waves. For BHs of masses $M_1$ and $M_2$, the binary capture cross section is given by~\cite{1989ApJ...343..725Q,Mouri:2002mc,Bird:2016dcv}:
\beq
\sigma_{\rm C} = \frac{ 2\pi}{M^4_{\rm P}} 
\left[    \bigg(\frac{85\pi}{6\sqrt{2}}\bigg)^{2}    \frac{(M_1+M_2)^{10} (M_1 M_2)^{2}}{ v^{18}}    \right]^{1/7} \simeq 
\frac{45}{v^{18/7}}   \frac{M^2}{M^4_{\rm P}} \biggr|_{M = M_1 = M_2},    
\eeq
where, in the last step, we have taken the $M_1 = M_2$ limit, which we will assume (for simplicity) throughout
this paper. 
Note that this cross section is relatively insensitive to the ratio of the BH masses,
 varying by only a factor of five across $M_1/M_2=10^{-3}$ to $10^3$ (for a fixed value of $M_1+M_2$).

A necessary, but not necessarily sufficient, condition for BHs to merge efficiently is that their binary capture rate 
exceeds the rate of Hubble expansion. The capture rate is given as follows: 
\be
\Gamma_{\rm C} = n_{\rm BH} \sigma_C v,
\ee
where $n_{\rm BH} = \rho_{\rm BH}/M$ is the BH number density and $v$ is their
relative velocity.
Using the general expression for $H$ in Eq.~(\ref{HTeff}) and $T_{\rm eff}$ in Eq.~(\ref{Teff}), the 
time-dependent ratio of these rates is 
\begin{eqnarray}
\label{capture-criterion}
\frac{\Gamma_{\rm C}}{H} \simeq      
45 \sqrt{  \frac{3}{8\pi \rho_{\rm T}} }
       \frac{M   }{  M^3_{\rm P}          }  \frac{   \rho_{\rm BH}  }{v^{11/7}}
\simeq 
92\fbh \frac{ M T_{\rm eff}^2}{M_{\rm P}^3 v^{11/7}} ,
\end{eqnarray}
where, in the last line, we have used
$\rho_{\rm BH} = f_{\rm BH} \rho_{\rm T}$ and write the total energy density in terms of the effective temperature. Note that both before and after BH domination, the capture rate decreases relative to Hubble as the universe expands, so 
if the initial conditions do not allow for capture, no mergers  will ever take place. 

Conversely, when $\Gamma_{\rm C} \gg H$, a typical BH will undergo many such encounters, potentially involving complex multi-body dynamics. Here, we simply assume that this continues until $\Gamma_{\rm C} \sim H$, at which point the process of binary capture effectively ceases, leaving the overwhelming majority of BHs in gravitationally bound binary systems; we refer to this as the time of  ``capture freeze-out''.  Setting $\Gamma_{\rm C} = H$ in Eq.~(\ref{capture-criterion}), the effective temperature at capture freeze-out
can be written
\begin{eqnarray}
\label{Tcf}
T_{\rm eff}(a_{\rm CF}) 
%&=& \bigg(\frac{16}{405\pi g_{\star}(T_{\rm eff})}\bigg)^{1/4} \frac{v^{11/14} M^{3/2}_{\rm P}}{M^{1/2}} \fbh(T_{\rm CF}) ^{-1/2} \\
\approx 2.6 \times 10^{9} \, {\rm GeV} \, \bigg(\frac{v}{10^{-3}}\bigg)^{11/14} \bigg(\frac{10^8 \, {\rm g}}{M_i}\bigg)^{1/2} \fbh(a_{\rm CF}) ^{-1/2},
\end{eqnarray}
where we have adopted a reference value of $v = 10^{-3}$, motivated by the gravitational acceleration associated with density perturbations in the early universe (see Appendices~\ref{App:velocity} and~\ref{hubbleflow}).
Note that Eq.~(\ref{Tcf}) simultaneously covers two possible scenarios:
\begin{itemize}
 \item{\bf Capture Freeze-Out in Radiation Domination:} If capture freeze-out occurs during radiation domination,  $\Gamma_{\rm C}/H \propto a^{-1}$,
  $T_{\rm eff}(a_{\rm CF}) = T_i/a_{\rm CF}$, and $f_{\rm BH}(a_{\rm CF}) = \fbhi a_{\rm CF}$, where the scale factor at capture freeze-out 
  can be obtained from Eq.~(\ref{capture-criterion}),
 \be
 \label{aCF}
 a_{\rm CF} \simeq  \frac{M_{\rm P}^3 v^{11/7}}{92 M \fbhi  T_i^2}~.
 \ee

\item{\bf Capture Freeze-Out in Black Hole Domination:} If capture freeze-out occurs during BH domination,
$\Gamma_{\rm C}/H \propto a^{-3/2}$ and $\fbh(a_{\rm CF}) =1$,
 so we can write $T_{\rm eff} = (30 \rho_{\rm BH}/\pi^2 g_\star)^{1/4}$, where $\rho_{\rm BH}$ is evaluated at capture freeze-out. 
% Note that because BH domination can follow a initially radiation dominated era with initial BH fraction $\fbhi$, for $a_{\rm CF} > a_{\rm D} = 1/\fbhi$, as discussed below Eq.~(\ref{Teff}).
\end{itemize}

% \approx  15 \bigg(\frac{M}{10^8 \, {\rm g}}\bigg) \, \bigg(\frac{T_{\rm eff}}{10^{10} \, {\rm GeV}}\bigg)^2 \, \bigg(\frac{10^{-3}}{v}\bigg)^{11/7} 
% \fbh(a),

%%%%%%%%%%%%%%%%%%%%%%%%

In addition to the binary capture mechanism described above, it is possible that binaries could be formed as a result of the tidal effects of the field of surrounding BHs (or cosmological density perturbations)~\cite{Nakamura:1997sm,Ioka:1998nz,Ali-Haimoud:2017rtz,Raidal:2018bbj,Garriga:2019vqu}.  In such a scenario, BHs start out in a quasi-stationary state, but become bound into pairs and decouple from the Hubble flow when the mutual gravitational binding energy of the pair exceeds the corresponding kinetic energy associated with the flow. If this occurs during an epoch of radiation domination, then binary formation occurs when the BHs' local contribution to the energy density exceeds the background radiation density.   In addition, weak tidal torques from the nearby field of surrounding BHs can lead to the acquisition of angular momentum by the close pair and prevent a direct collision. Although this mechanism naively leads to the formation of binaries with highly eccentric ($1-e \ll 1$) orbits, subsequent interactions with later infalling BHs can significantly reduce the eccentricity, semi-circularizing the binary orbit, or even leading to the disruption of the binary~\cite{Raidal:2018bbj}.

%A small numerical simulation of this process involving 70 BH demonstrated this effect could be significant, with most binaries being disrupted and the remaining binaries having much decreased eccentricity \cite{Raidal:2018bbj}.  If binary capture is efficient only BH binaries formed in relatively rare under-dense regions are free from this post-formation re-processing of their orbital parameters.   A full analysis of this possibility including the subsequent infall and clustering effects requires a dedicated numerical simulation, going beyond the scope of this work.  Thus in our discussion of the typical binary separation and in-spiral time in the subsequent section we introduce a parameter $\lambda$ to cover this possible binary formation mechanism and parameterise the effect of later infall circularisation on both mechanisms.

%%%%%%%%%%%%%%%%%%%%%%%%%%
% 		 Subsection: Binary Separation Distance
%%%%%%%%%%%%%%%%%%%%%%%%%%
  
\subsection{Binary Separation Distance}

Once the capture rate freezes out, the binaries begin the process of inspiral. The timescale for this process depends on the initial separation distance between the BHs in a bound pair, $L_{\rm CF}$, which is bounded from above as follows:
\be
\label{LCF}
L_{\rm CF} < n_{\rm CF}^{-1/3} = \left(    \frac{M}{ \rho_{\rm BH}(a_{\rm CF}) }   \right)^{1/3},
\ee
where $n_{\rm CF}$ is the number density of BHs at the time of capture freeze-out. One can expect the typical value of $L_{\rm CF}$ to be smaller than this upper limit, however. And realistically, there will be a distribution in the binary orbital parameters of eccentricity and semi-major axis. During the period with $\Gamma_C \gg H$,  a typical BH will experience multiple strong gravitational encounters, which may alter the orbital parameters of a binary in various ways, leaving it more or less tightly bound, or even disrupted. For example, for binaries with more gravitational binding energy than the average BH's kinetic energy, 3-body encounters will statistically tend to increase the binding energy of the binary~\cite{Heggie:1975tg}, thereby shortening this distance by a potentially important amount.

Since modeling these complicated dynamics is beyond the scope of this work, we will proceed by describing these orbits in terms of a characteristic initial separation for the case of a circular orbit. We parameterize our ignorance by introducing the parameter $\lambda$, such that the typical BH binary separation distance at the time of capture freeze-out is given as follows:
\begin{equation}
L_{\rm CF} = \lambda n_{\rm CF}^{-1/3}.
\end{equation}

A single radiative capture event between two isolated BHs typically results in a highly eccentric ($e\sim 1$) orbit~\cite{Cholis:2016kqi}. For equal semi-major axes, BHs in eccentric orbits inspiral much more quickly than those in circular ($e=0$) orbits (see below, in Eq.~\ref{tinsp}), by a factor of approximately $3.6 \, (1-e^2)^{7/2}$~\cite{Peters:1964zz}. Note, however, that by choosing $\lambda$ appropriately, one can mimic any value of the eccentricity. For example, $e\sim 0.9$ ($\sim 0.99$) can be absorbed by decreasing $\lambda$ by a factor of $\sim 3$ ($\sim 20$). The parameter $\lambda$ also allows us to account for any effects associated with the binary formation through tidal forces, as discussed above. Also note that one expects highly eccentric orbits to be more prone to be disrupted and circularized by gravitational encounters and tidal forces, as demonstrated by the 70-body simulations described in Ref.~\cite{Raidal:2018bbj}. In the calculations that follow, we will adopt a benchmark value of $\lambda=0.1$, but keep in mind that this parameter will depend on the duration and other details of the period in which $\Gamma_{\rm C} \gsim H$.

\subsection{Inspiral Timescale}

Once a binary system has formed, the BHs will gradually inspiral toward one another. Assuming that gravitational wave emission dominates this process, the inspiral time $t_{\rm I}$ can be written as~\cite{Peters:1964zz}: 
\begin{eqnarray}
\label{tinsp}
t_{\rm I} =  \frac{5 M^6_{\rm P}}{512  M^3} \frac{\lambda^4}{n^{4/3}_{\rm BH}(a_{\rm CF})} 
 \approx  1.4\times 10^{-21} \, {\rm s} \, \bigg(\frac{M}{10^8 \, {\rm g}}\bigg) \bigg(\frac{\lambda}{0.1}\bigg)^4 \bigg(\frac{10^{-3}}{v}\bigg)^{88/21} \fbh(a_{\rm CF})^{4/3},
\end{eqnarray}
where in the last step we have  rewritten $n_{\rm BH}(a_{\rm CF})$ as follows:
\begin{equation}
n_{\rm BH}(a_{\rm CF}) = \frac{\rho_{\rm BH} (a_{\rm CF})}{M} = \frac{\pi^2 g_{\star} T^4_{\rm eff}(a_{\rm CF})f_{\rm BH}(a_{\rm CF})}{30 M}  ,
\end{equation}
and then used Eq.~(\ref{Tcf}) to relate this to $M$, $v$ and $f_{\rm BH}(a_{\rm CF})$.

In order for a merger to take place, the inspiral time must be shorter than the Hawking evaporation lifetime ($t_{\rm I} \lsim \tau$). The ratio of these timescales is given by: 
\be
\label{tinsp-tevap}
\frac{t_{\rm I}}{\tau} = 
\left( \frac{5 M^6_{\rm P}}{512  M^3} \frac{\lambda^4}{n^{4/3}_{\rm CF}} \right) 
\left( \frac{3 M^4_{\rm P}}{  \langle \ell^{-1} \rangle    M^3 } \right) 
\sim  \bigg(\frac{\lambda}{0.1}\bigg)^4 \bigg(\frac{10^{-3}}{v}\bigg)^{88/21} \bigg(\frac{0.2 \, {\rm g}}{M}\bigg)^2 
   \bigg(\frac{235}{\langle \ell^{-1}\rangle}\bigg) \,  \fbh(a_{\rm CF})^{4/3}  ,
\ee
where, again, $\fbh(a_{\rm CF})$ is the BH fraction at capture freeze-out. Demanding that the merger take place 
before evaporation implies the following condition on the population mass: 
\begin{eqnarray}
M \gsim 0.2 \, {\rm g }\,  \bigg(\frac{235}{\langle \ell^{-1}\rangle}\bigg)^{1/2} \bigg(\frac{\lambda}{0.1}\bigg)^2 \bigg(\frac{10^{-3}}{v}\bigg)^{44/21} \fbh(a_{\rm CF})^{2/3}.
\end{eqnarray}
This condition simultaneously accounts for capture freeze-out during either radiation or BH domination.
For BHs that satisfy the above conditions, binary formation and merger are expected to occur prior to evaporation, leading to the production of gravitational waves as discussed in the following section.

%%%%%%%%%%%%%%%%%%%%%%%%%%
%%%%%%%%%%%%%%%%%%%%%%%%%%

% 		 Figure: Merger Neff with Teff

%%%%%%%%%%%%%%%%%%%%%%%%%%
%%%%%%%%%%%%%%%%%%%%%%%%%%  

\section{The Contribution to $N_{\rm eff}$ From Gravitational Waves}
\label{Sec:NeffGW}

When BHs merge, they emit an energy equal to a fraction $\xi \sim {\mathcal O}(10\%)$ of their mass into gravitational waves. From that point onward, the evolution of the energy densities in BHs, radiation and gravitational waves evolve as follows:
\begin{eqnarray}
\dot \rho_{\rm BH} = -3\rho_{\rm BH} H + \rho_{\rm BH}  \frac{ \dot M }{M},~~~
\dot \rho_{R}  = -4\rho_{R} H - \rho_{\rm BH}  \frac{ \dot M }{M},~~~
\dot \rho_{\rm GW}  = -4\rho_{\rm GW} H, 
\end{eqnarray}
where the Hubble rate is determined using $\rho_{\rm T} = \rho_{\rm BH}+\rho_{R}+\rho_{\rm GW}$. The energy density in gravitational waves evolves as radiation, and contributes to the measured value of the effective number of neutrino species, $N_{\rm eff}$.

 The contribution of these gravitational waves to $N_{\rm eff}$ is related to their fraction of the total energy density at the time of matter-radiation equality, $t_{\rm EQ}$. After accounting for SM entropy dumps, this is related as follows to the fraction of the energy density immediately after BH evaporation (see Ref.~\cite{Hooper:2019gtx} for a derivation):
\begin{eqnarray}
\label{rho-ratio}
\frac{\rho_{\rm GW}(t_{\rm EQ})}{\rho_{R}(t_{\rm EQ})} &=& \frac{\rho_{\rm GW}(\tau)}{\rho_{R}(\tau)}   \left(       \frac{g_*(T_{\rm evap})}{g_*(T_{\rm EQ})} \right)    \left( \frac{g_{*S}(T_{\rm EQ})}{g_{*S}(T_{\rm evap})} \right)^{4/3},
    \end{eqnarray}
where $T_{\rm EQ} \simeq 0.75$ eV and $T_{\rm evap}$ is the temperature of the SM bath immediately after BH evaporation has ended.

\begin{figure}[t]
\includegraphics[width=0.45\textwidth]{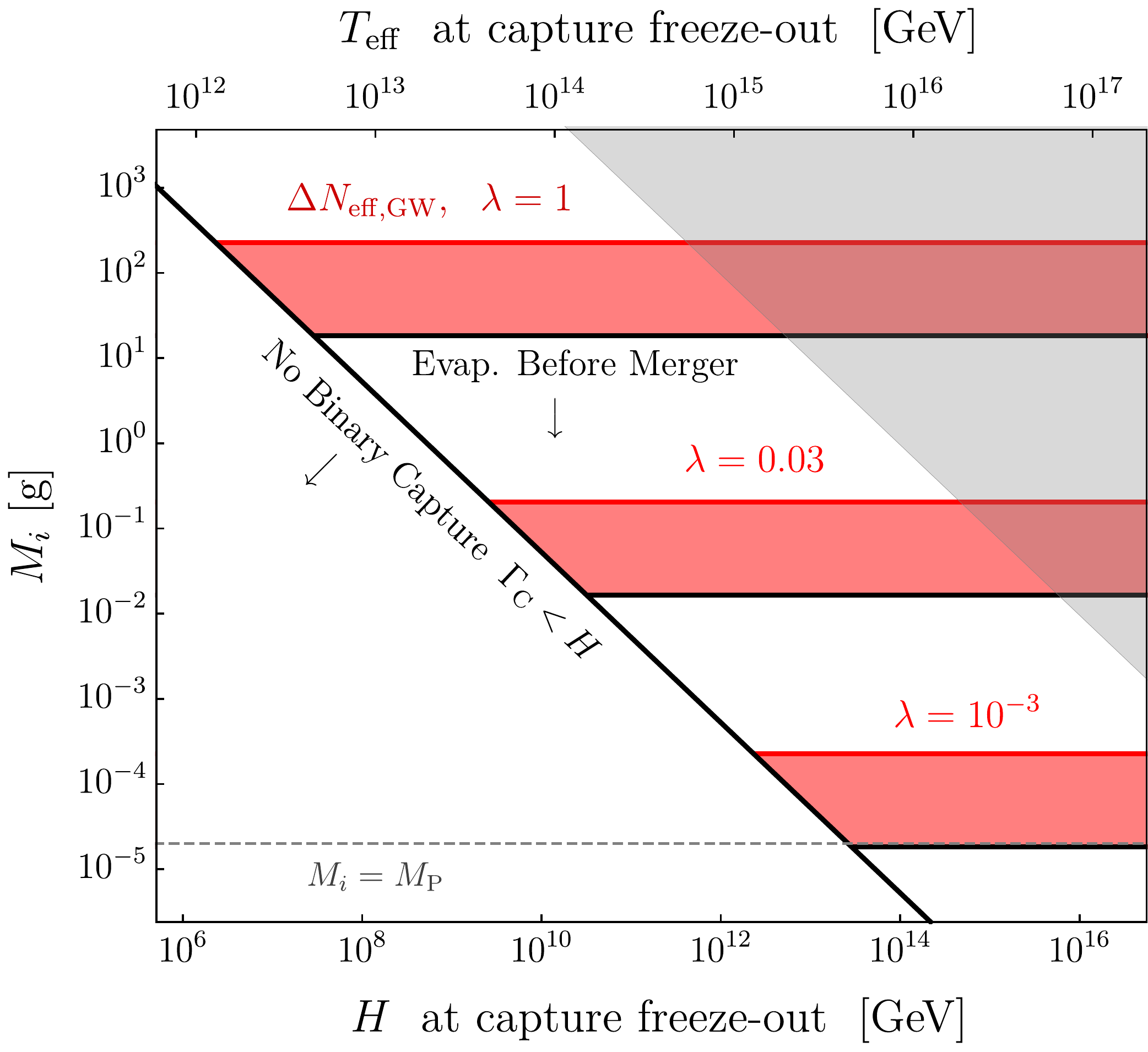} ~~
\includegraphics[width=0.45\textwidth]{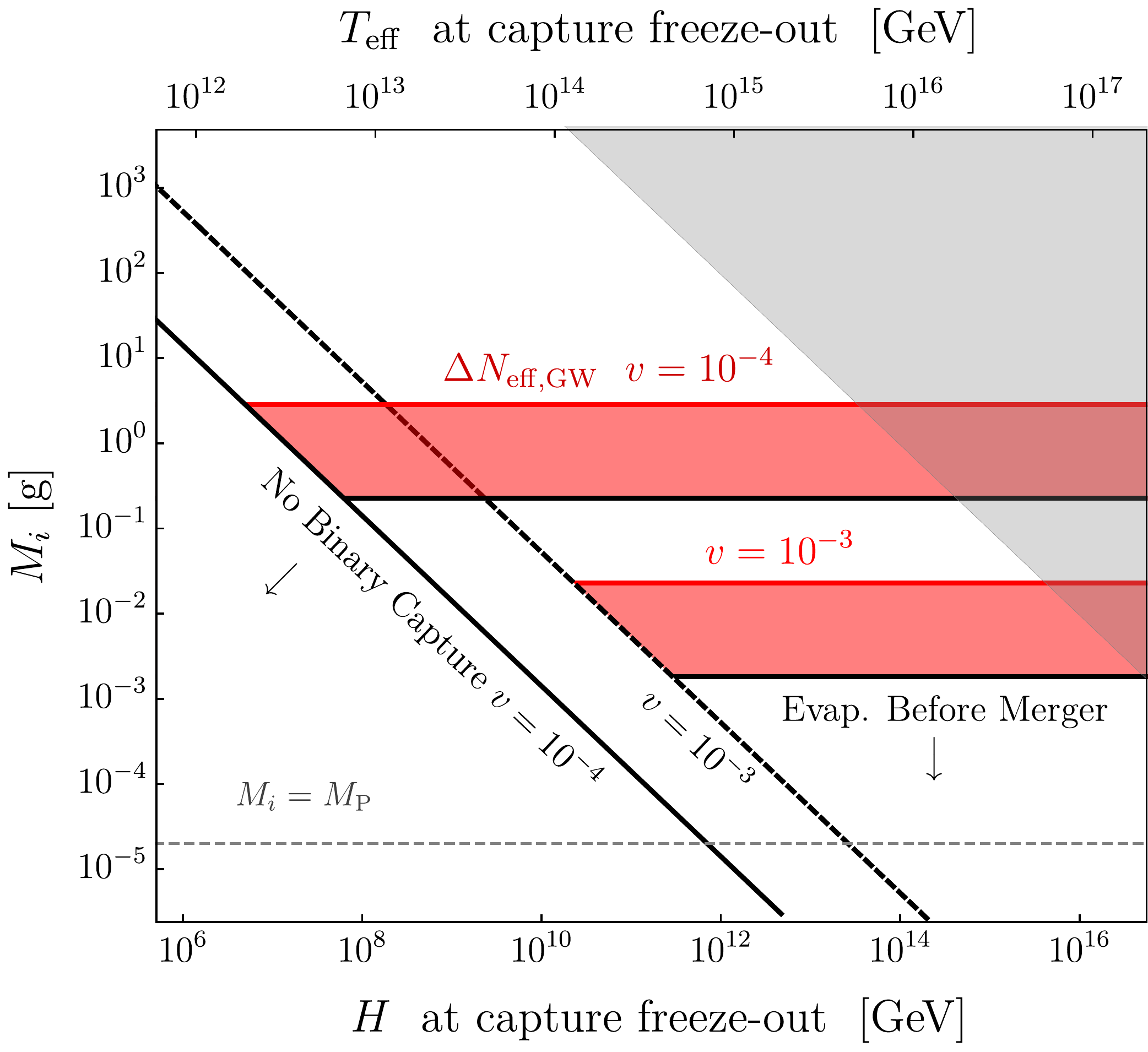} 
\caption{Regions of parameter space in which the gravitational waves from BH mergers in the early universe constitute an energy density equivalent to $\Delta N_{\rm eff} = 0.01-0.3$ (red bands; see Eq.~\ref{resultGW}), 
assuming the universe is BH dominated at the time of capture freeze-out. This range of $\Delta N_{\rm eff}$ is both consistent with current constraints and within the projected reach of upcoming CMB observations \cite{Abazajian:2016yjj}. We show these results in terms of the initial BH mass, $M_i$, and the effective temperature, $T_{\rm eff}$, or Hubble rate, $H$, as evaluated at the time of capture freeze-out.
{\bf Left:} $\Delta N_{\rm eff} = 0.01-0.3$ bands for $v=10^{-3}$ and three choices of $\lambda$, which parametrizes our ignorance of the BH
orbital parameters at the time of capture freeze-out. To the left of the black diagonal line, efficient binary capture never occurs. {\bf Right:} As in the left panel, but for $\lambda=0.01$ and two values of the BH velocity, $v$. The grey shaded portions of these figures represent unphysical regions of parameter space, in which many of the BHs are expected to overlap ($2r_{\rm Sch} > n_{\rm BH}^{-1/3}$).}
\end{figure}

If the universe is radiation dominated at $t = \tau$, then $T_{\rm evap} = T_i/a(\tau) = T_i (t_{\rm i}/\tau)^{1/2}$. 
If, on the other hand, BHs dominate the energy density of the universe at $t = \tau$, then their Hawking evaporation
is responsible for reheating the universe and setting $\rho_{R}(\tau)$ in Eq.~(\ref{rho-ratio}). 
Assuming nearly instantaneous BH evaporation near $t \sim \tau$, we can estimate $T_{\rm evap}$ according to 
\be
\rho_{\rm BH}(\tau) \approx  \frac{\pi^2}{30} g_\star(T_{\rm evap}) T_{\rm evap}^4~~~~\text{(Assuming BH Domination)}.
\ee
Furthermore, if the BH binaries merge during BH domination, we can rewrite 
Eq.~(\ref{rho-ratio}) as 
\begin{eqnarray}
\label{ratio}
\frac{\rho_{\rm GW}(t_{\rm EQ})}{\rho_{R}(t_{\rm EQ})} &=& \frac{\rho_{\rm GW}(\tau)}{\rho_{\rm BH}(\tau)}   \left(       \frac{g_*(T_{\rm evap})}{g_*(T_{\rm EQ})} \right)    \left( \frac{g_{*S}(T_{\rm EQ})}{g_{*S}(T_{\rm evap})} \right)^{4/3} \\
&=& \xi \, \bigg(\frac{t_{\rm I}}{\tau}\bigg)^{2/3}   \left(       \frac{g_*(T_{\rm evap})}{g_*(T_{\rm EQ})} \right)    \left( \frac{g_{*S}(T_{\rm EQ})}{g_{*S}(T_{\rm evap})} \right)^{4/3},
\end{eqnarray}
where we first blueshifted $\rho_{\rm GW}(\tau)$ from $\tau \to t_{\rm I}$ and then
used $\rho_{\rm BH} \propto H^2 = (2/3t)^2$ to rewrite $\rho_{\rm GW}(t_{\rm I})/\rho_{\rm R}(\tau)$.
Finally, using Eq.~(\ref{tinsp}) to evaluate the ratio $t_{\rm I}/\tau$, the contribution to $N_{\rm eff}$ from gravitational waves can be related to the ratio of energy densities in Eq.~(\ref{ratio}) as follows:
\be
\label{NeffGW}
\Delta N_{\rm eff, GW} = \frac{\rho_{\rm GW}(t_{\rm EQ})}{\rho_{\rm R}(t_{\rm EQ})}  \bigg[\frac{8}{7}\bigg(\frac{11}{4}\bigg)^{4/3}
\!\!
+N_{\nu}\bigg] 
= \xi \, \bigg(\frac{t_{\rm I}}{\tau}\bigg)^{2/3} \!   \left(       \frac{g_*(T_{\rm evap})}{g_*(T_{\rm EQ})} \right)    \left( \frac{g_{*S}(T_{\rm EQ})}{g_{*S}(T_{\rm evap})} \right)^{4/3}   \bigg[\frac{8}{7}\bigg(\frac{11}{4}\bigg)^{4/3}+N_{\nu}\bigg] ,
\ee
where $N_{\nu} \simeq 3.046$  is the SM contribution to $N_{\rm eff}$. 
Assuming $T_{\rm evap} \gg m_t \approx 175$ GeV, Hawking radiation will produce the full SM particle spectrum as radiation
 upon evaporation, so Eq.~(\ref{NeffGW}) becomes
\be
\label{resultGW}
 \Delta N_{\rm eff, GW}  \approx 0.3 \times \bigg(\frac{\xi}{0.1}\bigg) \, \bigg(\frac{195}{\langle \ell^{-1}\rangle}\bigg)^{2/3} \!\!\! \bigg(\frac{\lambda}{0.1}\bigg)^{8/3} \bigg(\frac{10^{-3}}{v}\bigg)^{176/63} \bigg(\frac{0.2 \, {\rm g}}{M}\bigg)^{4/3}
\fbh(a_{\rm CF})^{8/9},
\ee
where, as before, $\fbh(a_{\rm CF})$ is the BH fraction at capture freeze-out and this expression captures the $\Delta N_{\rm eff}$ contribution
regardless of whether capture freeze-out occurs before or after BH domination.

 One should keep in mind that these expressions are valid only in the case that $t_{\rm I} < \tau$ (otherwise no mergers, and thus no gravitational waves, will be produced). Therefore the maximum contribution is given by $\Delta N_{\rm eff, GW} \approx 0.3 \times (\xi/0.1)$, close to the current upper limit on this quantity~\cite{Aghanim:2018eyx}. We remind the reader that we have treated the BH evaporation as instantaneous in our calculation. More realistically, the evaporation will take place somewhat gradually, and will be delayed by the merger (which increases the BH mass and thus the evaporation time). 

Although the results presented in this section are consistent with existing constraints from measurements of the CMB~\cite{Aghanim:2018eyx}, the predicted contribution is potentially within the projected reach of stage IV CMB experiments, $\Delta N_{{\rm eff}} \sim 0.02$~\cite{Abazajian:2016yjj,Baumann:2017gkg,Hanany:2019lle}. In particular, this calculation demonstrates that, in order for these gravitational waves to produce a potentially measurable contribution to $N_{\rm eff}$, the BHs must merge not long before they evaporate (\ie~$t_{I}$ must be comparable to, but not greater than, $\tau$). This is because the energy density of gravitational waves evolves as $\rho_{\rm GW} \propto a^{-4}$ during this time, while $\rho_{\rm BH} \propto a^{-3}$, causing the fraction of the total energy density in gravitational waves to decrease. Once the BHs evaporate and the universe is dominated by radiation, however, the fraction of the energy density in gravitational waves remains constant (up to entropy dumps associated with the changing value of $g_{\star}(T)$).   
  
These results are particularly interesting in light of the $4.4\sigma$ discrepancy between the value of the Hubble constant as determined from local measurements~\cite{Riess:2019cxk,Riess:2018byc,Riess:2016jrr} and as inferred from the temperature anisotropies of the CMB~\cite{Aghanim:2018eyx}. This tension can be substantially relaxed if $\Delta N_{\rm eff} \sim 0.1-0.3$~\cite{Bernal:2016gxb,Aylor:2018drw,Weinberg:2013kea,Shakya:2016oxf,Berlin:2018ztp,DEramo:2018vss,Dessert:2018khu,Escudero:2019gzq}. From this perspective, a population of primordial BHs that merge shortly before evaporating can provide an attractive way of addressing the Hubble tension.

 \begin{figure}[t]
 \label{NeffRD}
\includegraphics[width=0.45\textwidth]{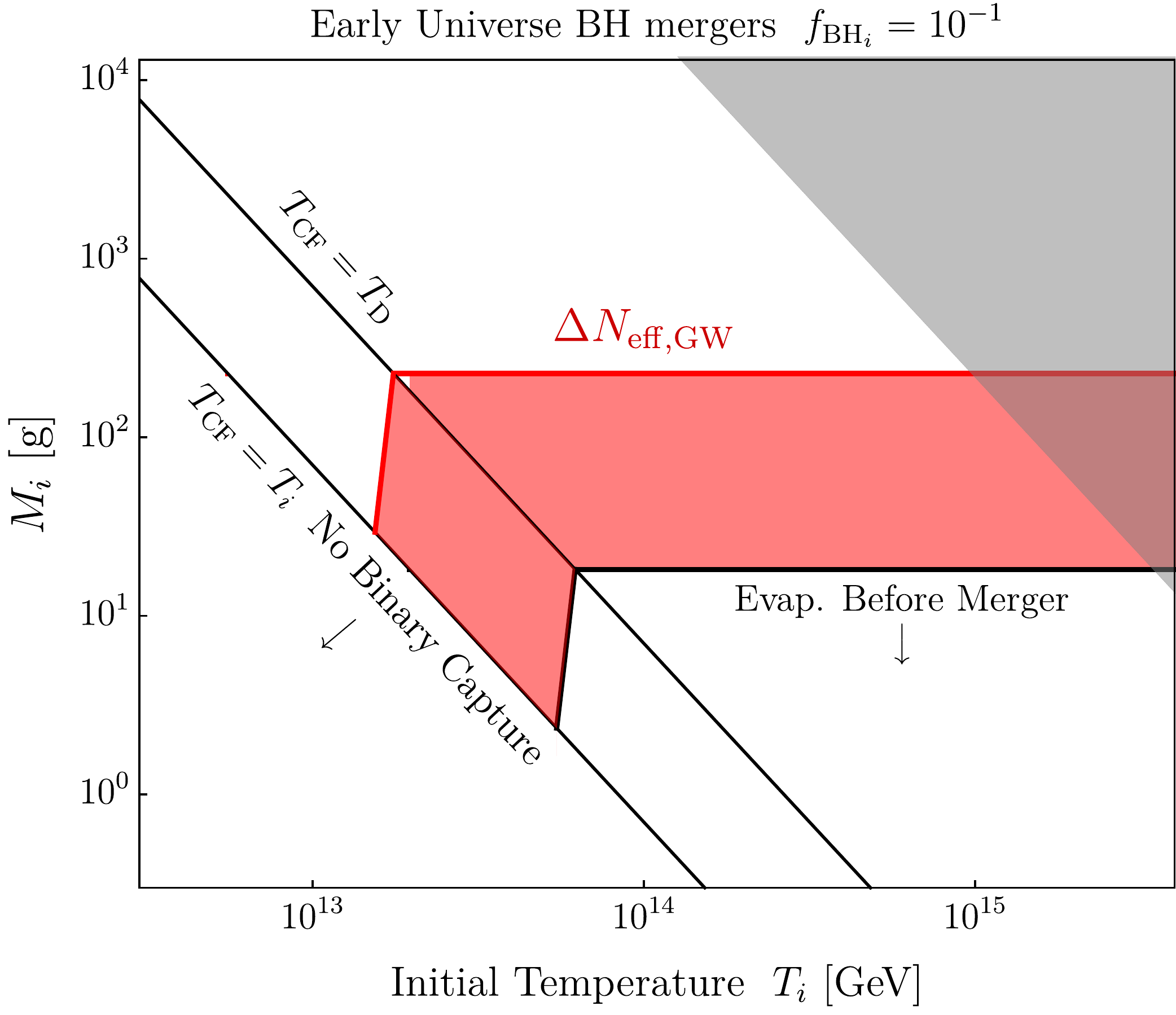} ~~
\includegraphics[width=0.45\textwidth]{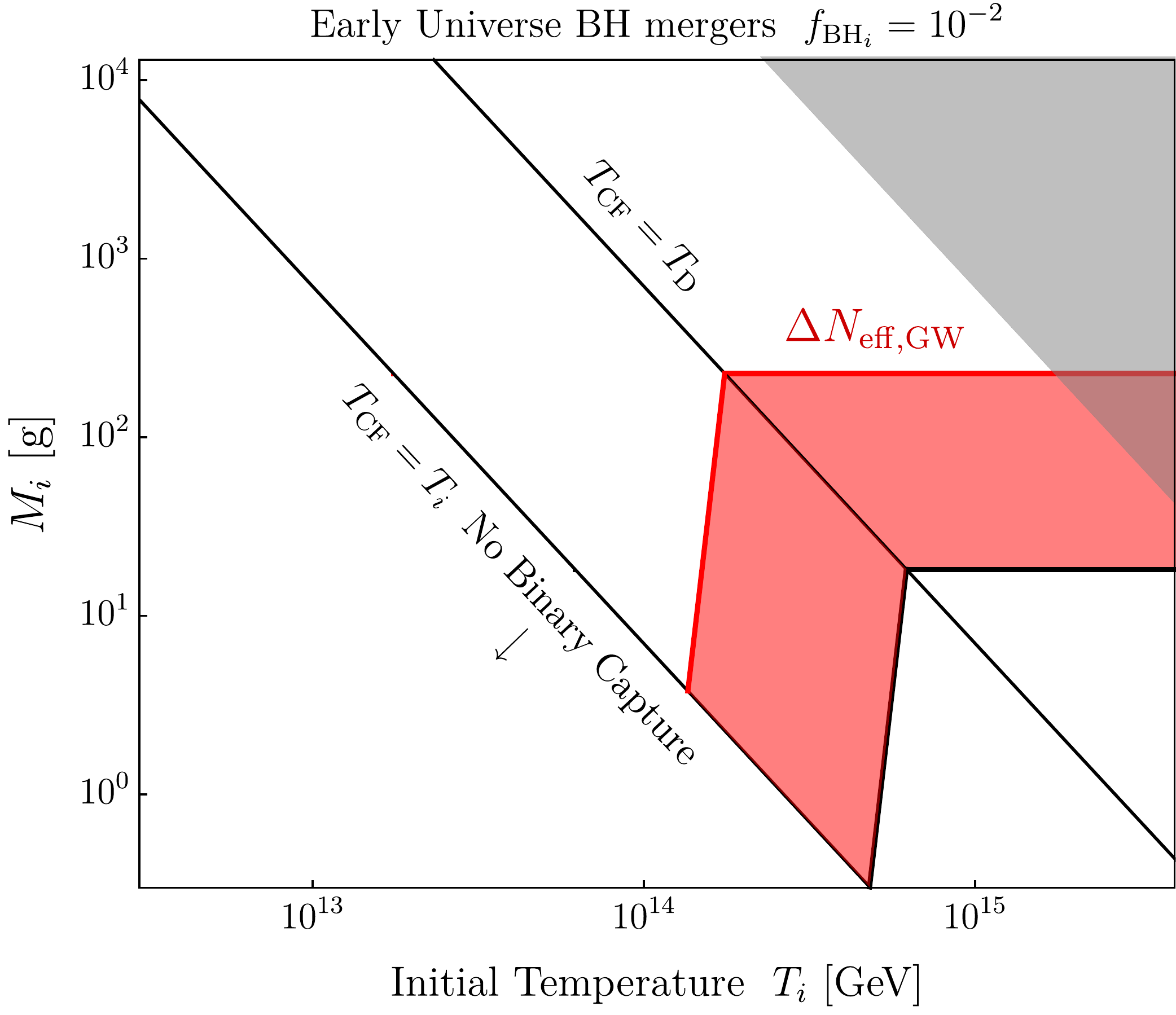} ~~
\caption{Regions of parameter space in which the gravitational waves from BH mergers in the early universe constitute an energy density equivalent to $\Delta N_{\rm eff} = 0.01-0.3$ (red bands; see Eq.~\ref{resultGW}), for initial conditions corresponding to a radiation-dominated state with temperature, $T_i$, and with a fraction, $f_{{\rm BH}_i}$, of the total energy density in black holes. This range of $\Delta N_{\rm eff}$ is both consistent with current constraints and within the projected reach of upcoming CMB observations \cite{Abazajian:2016yjj}. To the left of the left-most black diagonal line, efficient binary capture never occurs. The right-most black diagonal line denotes the parameter space where BH domination occurs at the same time of capture freeze out ($T_{\rm CF}=T_{\rm D}$). In each frame we had adopted $v=10^{-3}$ and $\lambda=1$. The grey shaded portions of these figures represent unphysical regions of parameter space, in which many of the BHs are expected to overlap ($2r_{\rm Sch} > n_{\rm BH}^{-1/3}$).}
\end{figure}

\section{The Contribution to $N_{\rm eff}$ From Hot Gravitons} 
\label{Sec:NeffHE}

In this section, we calculate the energy density in the gravitons that are emitted as Hawking radiation from a population of rotating BHs. In particular, we will consider a population of BHs with a distribution of spins as described in Ref.~\cite{Fishbach:2017dwv}, which is appropriate for any scenario in which most of the BHs have undergone at least one merger. As discussed in Sec.~\ref{Sec:HE}, such a population will emit approximately $f_{\rm G} \approx 0.47\%$ of their energy in the form of gravitons, a fraction that is significantly larger than is predicted in the case of non-rotating BHs.

As in the previous section, the contribution from gravitons to $\Delta N_{\rm eff}$ can be related to the energy density in gravitons at the time of equality \cite{Hooper:2019gtx}:
\be
\Delta N_{\rm eff, G} &=&  \frac{\rho_{\rm G}(T_{\rm EQ})}{\rho_{\rm R}(T_{\rm EQ})} \bigg[N_{\nu} + \frac{8}{7} \bigg(\frac{11}{4}\bigg)^{4/3}\bigg] 
=
 f_{\rm G} \, \bigg(\frac{g_{\star S}(T_{\rm EQ})}{g_{\star S}(T_{\rm evap})}\bigg)^{1/3} \bigg(\frac{g_{\star S}(T_{\rm EQ})}{g_{\star}(T_{\rm EQ})}\bigg) \bigg[N_{\nu}+\frac{8}{7}\bigg(\frac{11}{4}\bigg)^{4/3}\bigg] .
 \ee
Assuming $T_{\rm evap} \gg m_t$, so that the full SM radiation bath is populated after evaporation, 
the energy density in Hawking radiated gravitons is
 \be
\Delta N_{\rm eff, G} \approx 0.013 \, \bigg(\frac{f_{\rm G}}{0.0047} \bigg) \bigg(\frac{106}{g_{\star}(T_{\rm evap})}\bigg)^{1/3}. 
\ee
Note that for BH masses below $M \lsim 10^6-10^7$ g, the temperature after evaporation is quite high and $g_{\star}(T_{\rm evap}) \simeq 10^2$, resulting in a contribution of $\Delta N_{{\rm eff}, G} \sim 0.01$, just below the reach of future CMB measurements. For $M \sim 10^8-10^9$ g, however, the BHs will reheat the universe only to a temperature of $T_{\rm RH} \sim 10$ MeV, corresponding to $g_{\star}(T_{\rm RH}) \simeq 10$. In this case, we predict $\Delta N_{{\rm eff}, G} \sim 0.03$, within the projected sensitivity of stage IV CMB experiments~\cite{Abazajian:2016yjj,Baumann:2017gkg,Hanany:2019lle}.

In contrast with other cosmic backgrounds (\eg~the CMB, or the cosmic neutrino background), this background of gravitons consists of relatively high-energy particles. In particular, after numerically integrating the deposition of Hawking radiation over the lifetime of the BHs and redshifting their energy to the present era, we find that the mean energy of these gravitons today is given by:
\begin{eqnarray}
\langle E_G \rangle \sim 1.5 \, {\rm keV}  \, \bigg(\frac{M}{10^8\,{\rm g}}\bigg)^{1/2} \bigg(\frac{\langle \ell^{-1}\rangle}{195}\bigg)^{1/2} \bigg(\frac{14}{g_{\star}(T_{\rm RH})}\bigg)^{1/12}.
\end{eqnarray}
Given that this is a factor of $\sim 2\times 10^6 \times (M/10^8 \, {\rm g})^{1/2}$ times higher than the mean energy of a CMB photon, we refer to these particles as the ``hot graviton background''.  It would be fascinating to consider the possibilities for the detection of this signal.

\begin{figure}[t]
\hspace{-0.55cm}
\includegraphics[width=1.0\textwidth]{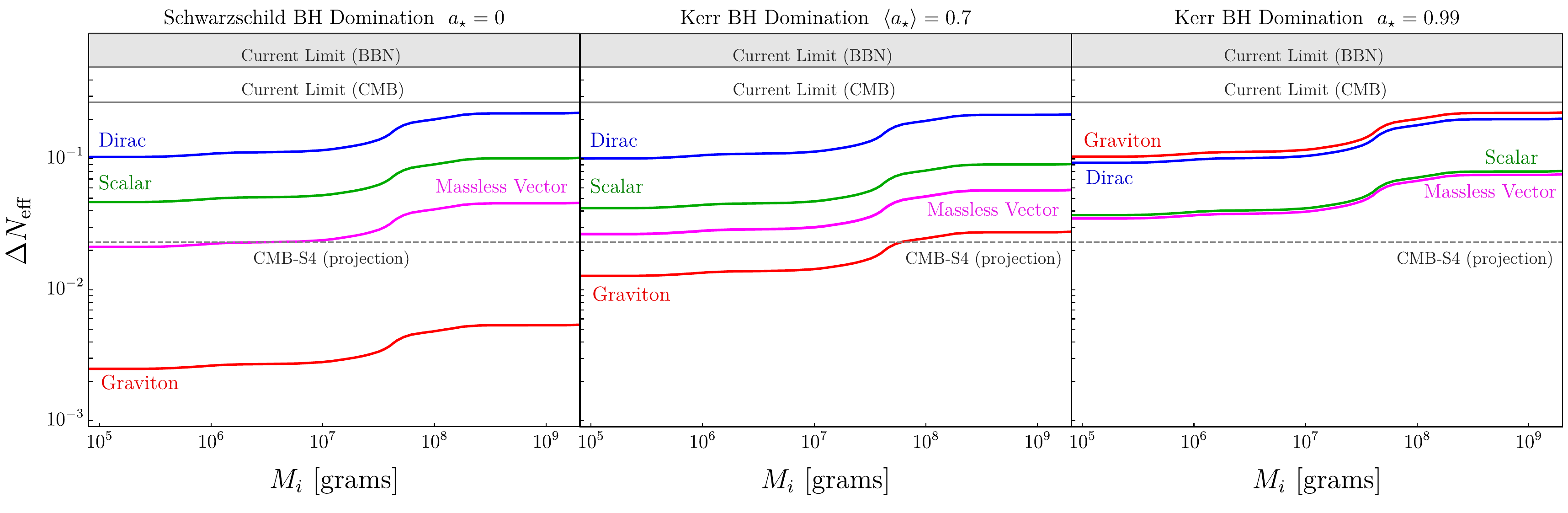} ~~~~
\caption{Contributions to $\Delta N_{\rm eff}$ from early universe Hawking evaporation, for a variety of hypothetical light and decoupled 
particle species, assuming Schwarzschild (left) and Kerr (center and right) BH domination at some point prior to BBN. Note that these plots do {\it not} assume that the universe was initially dominated by BHs, but only that BH domination occured {\it at some point} between
inflation and BBN. The left panel is 
adapted from Ref.~\cite{Hooper:2019gtx} and the Kerr contributions (center and right panels) is one of the main results of this paper. In the center frame, we adopt a distribution of angular momenta that peaks at $a_{\star} \sim 0.7$, as predicted for BHs that have undergone a merger~\cite{Fishbach:2017dwv}. In the right frame, we show results for the case of a population of BHs with very large initial spins, $a_{\star} = 0.99$. In each case, the contribution from a single massive vector is the sum of the contributions from a scalar and a massless vector.}
\label{HawkingNeff}
\end{figure}

Thus far, we have limited our discussion to Hawking evaporation into SM particles and gravitons. If there exist any other light and decoupled species, these too would be produced and contribute to the energy density of dark radiation. In Fig.~\ref{HawkingNeff}, we show this contribution to $\Delta N_{\rm eff}$ for various hypothetical light and decoupled particle species, assuming the that early universe had a BH dominated era. We show results for the case of non-rotating BHs (left)~\cite{Hooper:2019gtx}, as well as for Kerr BHs (center and right).

\section{The Gravitational Wave Spectrum} 
\label{Sec:GW}

Gravitational waves are generated throughout the stages of inspiral, merger and ring-down, producing the following respective contributions~\cite{Flanagan:1997sx,Ajith:2009bn,Cholis:2016xvo}:
%%
%\begin{eqnarray}
%\frac{dE}{df} \bigg|_{\rm inspiral} &=&  \frac{1}{3} \bigg(\pi^2 G^2\bigg)^{1/3} \, \frac{1}{f^{1/3}} \,\frac{M_1 M_2 }{(M_1+M_2)^{1/3}}, \,\,\,\,\,\,\,\,\,\,\,\,\,\,\,\,\,\,\,\,\,\,\, f < f_{\rm M},  \\
%%
%\frac{dE}{df} \bigg|_{\rm M} &=& \frac{1}{3} \bigg(\pi^2 G^2\bigg)^{1/3} \frac{f^{2/3}}{f_{\rm M}} \frac{M_1 M_2 }{(M_1+M_2)^{1/3}}, \,\,\,\,\,\,\,\, f_{\rm M}< f < f_{\rm RD},  \nonumber \\
%%
%\frac{dE}{df} \bigg|_{\rm RD} \,\,\,\,\,\,\, &=& \frac{1}{3} \bigg(\pi^2 G^2\bigg)^{1/3} \frac{f^2}{f_{\rm M} f^{4/3}_{\rm RD} \big[1+59.2 [(f/f_{\rm RD})-1]^2 \big]^2}  \frac{M_1 M_2 }{(M_1+M_2)^{1/3}}, \,\,\,\,\,\,\,\, f_{\rm RD}< f, \nonumber
%\end{eqnarray}
%%
%
\beq
\frac{dE}{df} =  \frac{1}{3} \pL \pi^2 G^2\pR^{1/3} \, \frac{M_1 M_2 }{(M_1+M_2)^{1/3}} \times
\begin{dcases}
f^{-1/3} &~~~~   f < f_{\rm M} \\
f^{2/3}/f_{\rm M} &~~~~   f_{\rm M}< f < f_{\rm RD} \\ 
\frac{f^2}{f_{\rm M} f^{4/3}_{\rm RD} \cbL[1+59.2 \bL(f/f_{\rm RD})-1\bR^2 \cbR^2}  &~~~~  f >f_{\rm RD}
\end{dcases}
\eeq
where $f_{\rm M} \simeq 8 \times 10^{36} \, {\rm Hz}  ({\rm g}/M_1+M_2)$ is the merger frequency and $f_{\rm RD} \simeq 4.5  f_{\rm M}$ is the ring down frequency. The spectrum is then redshifted from the time of the mergers to the present epoch, by a factor of $1+z = (T_{\rm evap}/T_{\rm CMB}) (g_{\star S}(T_{\rm evap})/g_{\star S}(T_{\rm CMB}))^{1/3}$, where $T_{\rm CMB}$ is the current temperature of the CMB. 

 \begin{figure}[t]
\includegraphics[width=0.5\textwidth]{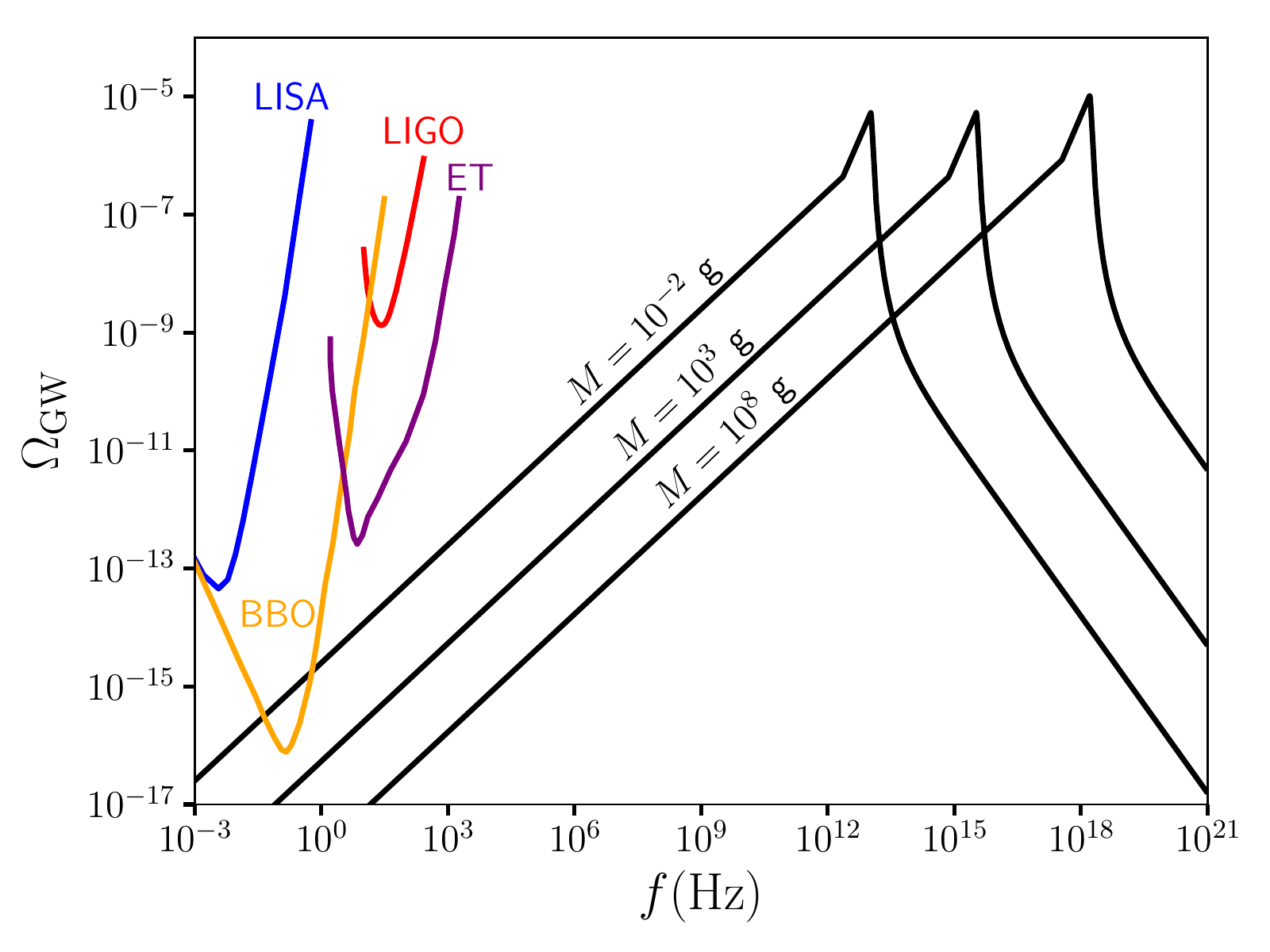} 
\caption{The spectrum of gravitational waves from BH mergers in the early universe, compared to the projected sensitivities of LIGO~\cite{TheLIGOScientific:2016dpb}, LISA~\cite{Bartolo:2016ami}, the Einstein Telescope (ET)~\cite{Punturo:2010zza,Hild:2010id}, and the Big Bang Observer (BBO)~\cite{Yagi:2011wg,Seto:2001qf}. These curves assume BH domination at the time of the mergers and that the mergers take place immediately prior to the BHs' evaporation. If the mergers take place well before evaporation, these curves should be appropriately redshifted (to lower frequencies) and suppressed by a factor of $(t_{\rm I}/\tau)^{2/3}$.}
\label{gwspec}
\end{figure}

In Fig.~\ref{gwspec}, we show the spectrum of the stochastic gravitational wave background from BH mergers in the early universe, assuming 1) BHs dominate the energy density at the time of their mergers, and 2) the mergers take place immediately prior to the BHs' evaporation. Departures from this second assumption result in a redshifting toward lower frequencies and an additional suppression by a factor of $(t_{\rm I}/\tau)^{2/3}$; see Eqn.~(\ref{NeffGW}). 
The predicted gravitational wave spectra are compared to the projected sensitivity of various gravitational wave experiments, including LIGO~\cite{TheLIGOScientific:2016dpb}, LISA~\cite{Bartolo:2016ami}, the Einstein Telescope (ET)~\cite{Punturo:2010zza,Hild:2010id}, and the Big Bang Observer (BBO)~\cite{Yagi:2011wg,Seto:2001qf}. In the most optimistic scenarios, we find that a future space-based experiment on the scale of BBO could potentially observe the stochastic background of gravitational waves predicted from BH mergers in the early universe.
 We present these results in terms of the quantity $\Omega_{\rm GW}$, which is related to $dE/df$ and the critical density as follows:
\begin{equation}
\Omega_{\rm GW} (f) =\frac{1}{\rho_c} \frac{d\rho_{\rm GW}}{d \ln f},
\end{equation}
where $\rho_c = (3 H^2_0 M^2_{\rm P} /8\pi )$ is  the present day critical density. 

For the gravitational wave spectra predicted in this class of scenarios, detectors optimized for their sensitivity to high-frequency gravitational waves may be promising~\cite{Ejlli:2019bqj,Arvanitaki:2012cn,Zheng:2018wjd,Goryachev:2014yra}. We leave further exploration of this topic to future work.

\section{Summary and Conclusions}
\label{Sec:conclusions}
 
If even a relatively small abundance of black holes was present in the early universe after inflation, the energy density of this population will be diluted more slowly than that of radiation. From this perspective, it is well-motivated to consider scenarios in which the early universe contained an era that was dominated by black holes. In order to ensure consistency with the measured light element abundances, such black holes must be light enough to evaporate prior to BBN. In previous work, it was pointed out that such black holes could potentially generate dark matter and dark radiation through the process of Hawking radiation. 
Dark radiation is not only a phenomenologically interesting but also an \emph{unavoidable} consequence of our scenario. Even without the existence of feebly-interacting light degrees-of-freedom beyond the Standard Model, there is a \emph{minimal} contribution to $\DNeff$ from the direct emission of gravitons. 
In this study we have expanded on past work to 
show how this contribution may be significantly enhanced in a very natural way.

In this paper, we have considered scenarios in which a large fraction of the black holes in the early universe became gravitationally bound into binary systems, and then merged prior to their evaporation. These mergers can produce a significant stochastic background of high-frequency gravitational waves, potentially within the reach of proposed space-based detectors. If these mergers take place only shortly before the black holes evaporate, the energy density of these gravitational waves could be as high as $\Delta N_{\rm eff} \sim 0.3$, potentially within the reach of next-generation CMB experiments.

Such mergers also leave the resulting black holes with significant angular momentum. As spinning black holes preferentially evaporate into high-spin particles, this class of scenarios can lead to the production of a significant background of hot ($\sim$\,eV-keV) gravitons in the universe today, with an energy density corresponding to $\Delta N_{\rm eff} \sim 0.01-0.03$. Other scenarios involving primordial black holes with near extremal angular momentum could produce an even larger energy density of energetic gravitons, up to $\Delta N_{\rm eff} \sim 0.3$ (see Fig.~\ref{HawkingNeff}).

% by BH spin. A natural way of producing rotating BHs from initially non-rotating ones is through mergers, the inspiral and coalscence dynamics of which contribute further to $\DNeff$ via gravitational waves.

\bigskip
\bigskip

\noindent{\it Note added:} While this work was being completed, we became aware of Ref. \cite{Inomata:2020lmk} which 
addresses a related subject. Our paper has some overlap with their discussion of the stochastic gravitational wave background
from early universe black hole mergers and our results are consistent with theirs. There is no overlap with our 
discussion here of dark radiation signals or Kerr black hole evaporation. 

\begin{acknowledgments}  
We would like to thank Camilo Garcia Cely for helpful comments. This manuscript has been authored by Fermi Research Alliance, LLC under Contract No. DE-AC02-07CH11359 with the U.S. Department of Energy, Office of High Energy Physics. 
The authors thank the Galileo Galilei Institute for hospitality during the Next Frontiers in the Search for Dark Matter workshop during
which this work was initiated. 
  RPB is supported by a joint Clarendon and Foley-Bejar Scholarship from the University of Oxford and Balliol College.

\end{acknowledgments}

\bibliography{PBH}

\appendix
\section{Black Hole Velocities Arising From Cosmological Perturbations}
\label{App:velocity}

Although the velocity distribution of a primordial BH population is model dependent, one can use the observed amplitude of density perturbations, $\delta \equiv \delta \rho/\bar \rho \sim 10^{-5}$, to motivate the typical speed of BHs in the early universe. In particular, even if a BH were produced at rest, it would be accelerated as a result of variations in the gravitational potential. We can estimate the kinetic energy acquired by a BH accelerated across a Hubble distance, $R_H \sim H^{-1}$, by an overdensity, $\delta$, in a Hubble volume $V_H = 4\pi R_H^3/3 = 4\pi/3H^{3}$ as
\begin{eqnarray}
 \frac{ v^2 }{2} &=& \frac{G  (\bar \rho V_H)  \delta }{R_H} \sim  \frac{4\pi G  \bar \rho   \delta}{3H^2} 
 \sim  \frac{  \delta}{2}.
\end{eqnarray}
Thus we expect a typical BH to be accelerated to a velocity on the order of $v\sim\sqrt{\delta} \sim 10^{-3}\mbox{--}10^{-2}$. Throughout this paper, we adopt a reference value of $v = 10^{-3}$, while acknowledging that higher values are possible. In particular, the amplitude of density perturbations on the relevant scales is currently not measured, and thus $\delta$ could be significantly larger than the value of $10^{-5}$ inferred from the CMB. We also note that density perturbations will continue to accelerate BHs throughout the early universe and thus will prevent the velocity of such objects from evolving with the scale factor, $\propto a^{-1}$, as might be naively expected.

 \section{Hubble Flow}
 \label{hubbleflow}
 
Throughout this study, we have assumed that BH binaries are not disrupted by Hubble expansion in the early universe. In this appendix, we explore this assumption and the criteria under which is it valid.
 
We begin by comparing the acceleration associated with Hubble expansion to that from gravitational attraction. In the Newtonian approximation, the distance between two BHs, $x$, evolves according to:
\begin{eqnarray}
\frac{d^2 x}{dt^2} &=& \bigg[\frac{dH}{dt} +H^2 \bigg] x - \frac{2MG}{x^2}, \\
&=&  \bigg[ \frac{4\pi G  \rho (1-3 w)}{3} \bigg] x - \frac{2MG}{x^2}, \nonumber
\end{eqnarray}
where $M$ is the mass of the BH, $\rho$ is the energy density and $w$ is the equation of state. During radiation domination (for which $1-3w=0$), the acceleration due to gravitational attraction always exceeds that associated with Hubble expansion. This is also the case during BH domination ($w=0$), so long as $\lambda \lsim (3/2\pi)^{1/3} \approx 0.8$. From this comparison, we find that any BHs that become close enough to one another to be gravitationally captured will be accelerated toward one another, so long as $\Gamma_{\rm C} \gsim H$ (see Sec.~\ref{Sec:mergers}) and $\lambda \lsim 0.8$.

\section{Black Hole Accretion}
\label{accretion}

A BH in a bath of isotropic radiation that is not undergoing background evolution will undergo Bondi-Hoyle accretion, gaining mass at the following rate~\cite{Bondi:1952ni}:
\beq \label{massacc1}
\frac{dM}{dt}\bigg|_{\rm Accretion} = \frac{4\pi \zeta M^2 \rho_{\rm R}}{M_{\rm P}^4(1+c_s^2)^{3/2}},   %+ \pL \frac{dM_{\rm BH}}{dt} \pR_{\rm Hawking},
\eeq
where $\zeta$ is an $\cO(1)$ constant and $c_s \simeq 1/\sqrt{3}$ is the sound speed in the radiation bath. Comparing this to the rate of Hawking evaporation in \Eq{rate}, we conclude that a BH in a static background will gain mass when $T_{\rm R} \gsim T_{\rm BH}$ and lose mass otherwise. Comparing the fractional accretion rate, $(1/M)dM/dt$, to the Hubble rate, we find that accretion may play a role at high temperatures, $T_{\rm R} \gsim 10^{12} \, {\rm GeV} \times (10^8 \, {\rm g}/M)^{1/2}$~\cite{Hooper:2019gtx}. However, this analysis of isotropic Bondi-Hoyle omits the effects of the background evolution. Accounting for cosmic expansion, only very anisotropic or unphysical fluids allow for a BH growth rate that matches or exceeds that of Hubble expansion~\cite{Carr:2010wk}.

\section{Runaway Mergers?}

In this study, we have assumed that BH mergers occur well after the end of the era of efficient binary capture. This ensures that each BH undergoes no more than one merger before evaporating. If this is not the case, BHs could be captured and inspiral many times in succession, potentially leading to runaway growth, and postponing evaporation until the era of BBN or beyond, where measurements of the light element abundances and the CMB provide very stringent constraints.

We can evaluate this possibility by comparing the inspiral time to the Hubble time, as evaluated at the time of capture freeze-out. We find that the inspiral time is shorter than the Hubble time at capture freeze-out only if the following condition is satisfied:
\begin{eqnarray}
f_{\rm BH}(a_{\rm CF})  &<& \frac{2^{30} \pi v^{55/7} }{3^{11}5^8 \lambda^{12}} 
\lsim 1.3\times 10^{-13} \bigg(\frac{v}{10^{-3}}\bigg)^{55/7} \bigg(\frac{0.1}{\lambda}\bigg)^{12},
\end{eqnarray}
which follows from combining Eqns.~(\ref{Tcf}) and~(\ref{tinsp}).

\end{document}